! TEX program = pdflatex
\documentclass[a4paper,11pt]{article}

\usepackage{setup}
\usepackage{jheppub}

\usepackage{cite}
\usepackage{ifthen}
\usepackage{subcaption}
\usepackage{siunitx}
\usepackage{enumitem}
\usepackage{booktabs}

\DeclareSIUnit\fb{\femto\barn}

\renewcommand{\tabcolsep}{1em}

\def\lapprox{\lower .7ex\hbox{$\;\stackrel{\textstyle <}{\sim}\;$}}
\def\gapprox{\lower .7ex\hbox{$\;\stackrel{\textstyle >}{\sim}\;$}}
\definecolor{lightgray}{HTML}{A6A39A}
\definecolor{darkgray}{HTML}{504E48}
\definecolor{silver}{HTML}{E0DFDE}
\definecolor{brown}{HTML}{5F4541}
\definecolor{beige}{HTML}{DCCCAC}
\definecolor{green}{HTML}{345F53}
\definecolor{yellow}{HTML}{F6B65A}
\definecolor{blue}{HTML}{568BCF}
\definecolor{red}{HTML}{AE1932}
\definecolor{orange}{HTML}{D16F15}

\makeatletter
\newcommand{\myitem}[1]{%
	\item[#1]\protected@edef\@currentlabel{#1}%
}
\makeatother

\preprint{{\raggedleft%
NIKHEF 2021-027 \\
BONN-TH-2021-10 \\
ZU-TH 51/21 \\
IPPP/21/40 \\
CERN-TH-2021-158 \\
}}

\title{Transverse momentum distributions in low-mass Drell-Yan lepton pair production at NNLO QCD}
\author[a,b]{R.~Gauld,}
\author[c,d]{A.~Gehrmann--De Ridder,}
\author[d]{T. Gehrmann,}
\author[e,f]{E.~W.~N.~Glover,}
\author[g]{A.~Huss,}
\author[c]{I.~Majer,}
\author[c]{A.~Rodriguez Garcia}

\affiliation[a]{Nikhef Theory Group, Science Park 105, 1098 XG Amsterdam, The Netherlands}
\affiliation[b]{Bethe Center for Theoretical Physics \& Physikalisches Institut der Universit\"at Bonn}
\affiliation[c]{Institute for Theoretical Physics, ETH, CH-8093 Z\"urich, Switzerland}
\affiliation[d]{Department of Physics, University of Z\"urich, CH-8057 Z\"urich, Switzerland}
\affiliation[e]{Institute for Particle Physics Phenomenology, Durham University,  Durham DH1 3LE, UK}
\affiliation[f]{Department of Physics, Durham University,  Durham DH1 3LE, UK}
\affiliation[g]{Theoretical Physics Department, CERN, CH-1211 Geneva 23, Switzerland}

\emailAdd{rgauld@uni-bonn.de}
\emailAdd{gehra@phys.ethz.ch}
\emailAdd{thomas.gehrmann@uzh.ch}
\emailAdd{e.w.n.glover@durham.ac.uk}
\emailAdd{alexander.huss@cern.ch}
\emailAdd{majeri@phys.ethz.ch}
\emailAdd{adrianro@phys.ethz.ch}

\abstract{The production of lepton pairs at low invariant mass and finite transverse momentum resolves QCD dynamics at the boundary between 
the perturbative and non-perturbative domains. We investigate the impact of NNLO QCD corrections on these observables 
at energies corresponding to the BNL RHIC collider and to 
fixed-target experiments.
Satisfactory perturbative convergence is observed in both cases. 
Only the collider data are found to be well-described by perturbative QCD, thus indicating the importance of non-perturbative effects in lepton-pair 
transverse momentum distributions at fixed target energies. \\
 }

\begin{document}
\maketitle
\flushbottom

\section{Introduction}
\label{intro}

The production of lepton  pairs in hadronic collisions (Drell-Yan process,~\cite{Drell:1970wh}) 
is mediated through a neutral electroweak gauge boson ($\gamma^*,Z$). 
It is a Standard Model benchmark process, which has been measured to very high precision at hadron colliders, thereby providing important insights 
into the structure of the colliding hadrons and into the dynamics of QCD
through inclusive measurements or transverse momentum distributions. Collider measurements of the Drell-Yan process focus 
mainly on lepton pairs with high pair invariant mass $Q$, including in particular the $Z$ resonance peak. The large mass scale $Q$ 
ensures that a reliable theory description of the Drell-Yan cross section and of associated transverse momentum distributions can be 
obtained within perturbative QCD using fixed-order predictions, combined with transverse momentum resummation if required.

The phenomenological importance of the Drell-Yan process has been driving precision calculations in particle theory, which 
are by now accomplished to next-to-next-to-leading order (NNLO) for the fully differential Drell-Yan process~\cite{Melnikov:2006kv,Catani:2009sm}.
Most recently, first results on the third order (N\textsuperscript{3}LO) corrections to the total Drell-Yan cross section~\cite{Duhr:2020seh} and its rapidity
distribution~\cite{Chen:2021vtu} were derived. The Born-level kinematics (order $\alpha_s^0$ in the QCD coupling constant) 
of the Drell-Yan process correspond to vanishing total transverse 
momentum $p_T$ of the lepton pair. Consequently, the Drell-Yan transverse momentum distribution receives its leading-order (LO) contribution 
at order $\alpha_s^1$. It is known to NNLO~\cite{Gehrmann-DeRidder:2015wbt,Gehrmann-DeRidder:2016cdi,Boughezal:2015ded}, 
which combines with transverse momentum resummation that has been 
derived~\cite{Bizon:2018foh,Ebert:2020dfc,Alioli:2021qbf,Camarda:2021ict,Ju:2021lah} to the third logarithmic order (N\textsuperscript{3}LL).

At lepton pair invariant masses well below the $Z$ resonance, virtual photon exchange largely dominates. With decreasing mass scale $Q$, 
the QCD coupling $\alpha_s$ increases, and the convergence of the perturbative expansion deteriorates. Low-mass Drell-Yan production 
can therefore probe the transition region between perturbative and non-perturbative QCD, and resolve aspects of the proton structure that go beyond 
the framework of collinear factorisation and parton distribution functions. 
Transverse momentum distributions of low-mass Drell-Yan pairs are particularly interesting in this context, since the 
simultaneous dependence on both $Q$ and $p_T$ enables to resolve the underlying dynamics in a multi-differential manner~\cite{Kang:2008wv}. 
Low-mass Drell-Yan production at moderate and low transverse momenta is experimentally challenging at high-energy colliders like the LHC or the Tevatron, since the resulting leptons emerge with transverse momenta that are too low to be cleanly detected (the low pile-up conditions at LHCb may provide an exception). 
Data on low-mass Drell-Yan production are mainly from fixed-target experiments, as well as from RHIC at BNL.

Extensive studies of the available measurements on transverse momentum distributions in low-mass Drell-Yan 
production has been performed recently~\cite{Bacchetta:2019tcu,BermudezMartinez:2020tys}, 
using next-to-leading order (NLO) QCD predictions~\cite{Melnikov:2006kv,Alwall:2014hca}
as baseline for the theoretical predictions, combined 
with next-to-leading logarithmic resummation. It was found in~\cite{Bacchetta:2019tcu,BermudezMartinez:2020tys} 
that these perturbative NLO QCD predictions were 
insufficient to describe the majority of available data, thereby yielding evidence for the relevance of non-perturbative effects such as intrinsic
partonic transverse momentum in the description of these observables. 
The low-mass Drell-Yan process has also been identified as ideal probe of transverse-momentum dependent parton distributions (TMDPDF), 
and first extractions of TMDPDFs for protons~\cite{Scimemi:2019cmh,Bertone:2019nxa,Bacchetta:2019sam}
 and pions~\cite{Vladimirov:2019bfa} have been performed on the available data sets.

This letter revisits the $p_T$ distributions of low-mass  Drell-Yan pairs in view of an improved perturbative description using for the first 
time NNLO QCD predictions. These are based on the calculation of NNLO QCD corrections to $Z$-boson production at large transverse 
momenta~\cite{Gehrmann-DeRidder:2016cdi}, which has been adapted to this kinematical situation. The calculation is implemented in 
the parton-level event generator \nnlojet, which uses the antenna subtraction
method~\cite{GehrmannDeRidder:2005cm,Daleo:2006xa,Currie:2013vh}
 to handle infrared singular real radiation at NNLO.

 For the numerical evaluations, we use the MMHT14 parton distribution functions~\cite{Harland-Lang:2014zoa}, with the associated values of $\alpha_s$. 
 We focus on two experimental data sets, which are sufficiently representative of the 
 full body of data on low-mass Drell-Yan production: from the PHENIX experiment~\cite{PHENIX:2018dwt} 
 at the BNL RHIC collider ($\sqrt{s}=200$~GeV) and from 
 the NuSea (E866) fixed-target experiment~\cite{NuSea:2003qoe,Webb:2003bj} at Fermilab ($\sqrt{s}=38.8$~GeV). 

The main objective of our work will be to verify the perturbative convergence of the 
$p_T$ distribution in low-mass Drell-Yan production, which is 
questionable~\cite{Bacchetta:2019tcu} in view of NLO corrections being of comparable size to the LO predictions. 
In probing extreme kinematics where the application of perturbative QCD starts to become questionable, our results will also 
allow to better quantify and constrain the amount of non-perturbative contributions in these observables, thereby enabling the future 
usage of the 
relevant data-sets in quantitative model studies of non-perturbative effects.

\section{Numerical results for Phenix}
\label{sec:results_phenix}

The PHENIX experiment is a multi-purpose detector at the BNL RHIC collider. It studies proton-proton and ion-ion collisions at various center-of-mass
energies.
For the measurement of low-mass lepton pair production~\cite{PHENIX:2018dwt},
 proton-proton collisions at a center-of-mass energy $\sqrt{s}=200$~GeV 
were analysed. The measurement 
of the transverse momentum ($p_T$) distribution 
of the lepton pair
in the range $p_T=0\text{--}6$~GeV was performed in a single bin in the lepton pair invariant mass $4.8~\GeV<Q<8.2~\GeV$, and
restricting the lepton pair rapidity to the forward and backward regions $1.2<|y|<2.2$, which are summed. 
The data were corrected to full acceptance for fiducial selection cuts on the individual leptons.

We compute the $p_T$-distribution using \nnlojet using the vector-boson-plus-jet calculation with the jet requirement
replaced by a minimum cut on $p_T$. Consequently, no fixed-order prediction is obtained for the leftmost bin, which includes 
$p_T=0$~GeV as boundary. The fixed-order prediction for the $p_T$ distribution diverges for low $p_T$, and should be supplemented by 
an all-order resummation of large logarithmic corrections. 
\begin{figure}[t]
    \includegraphics[width=\columnwidth]{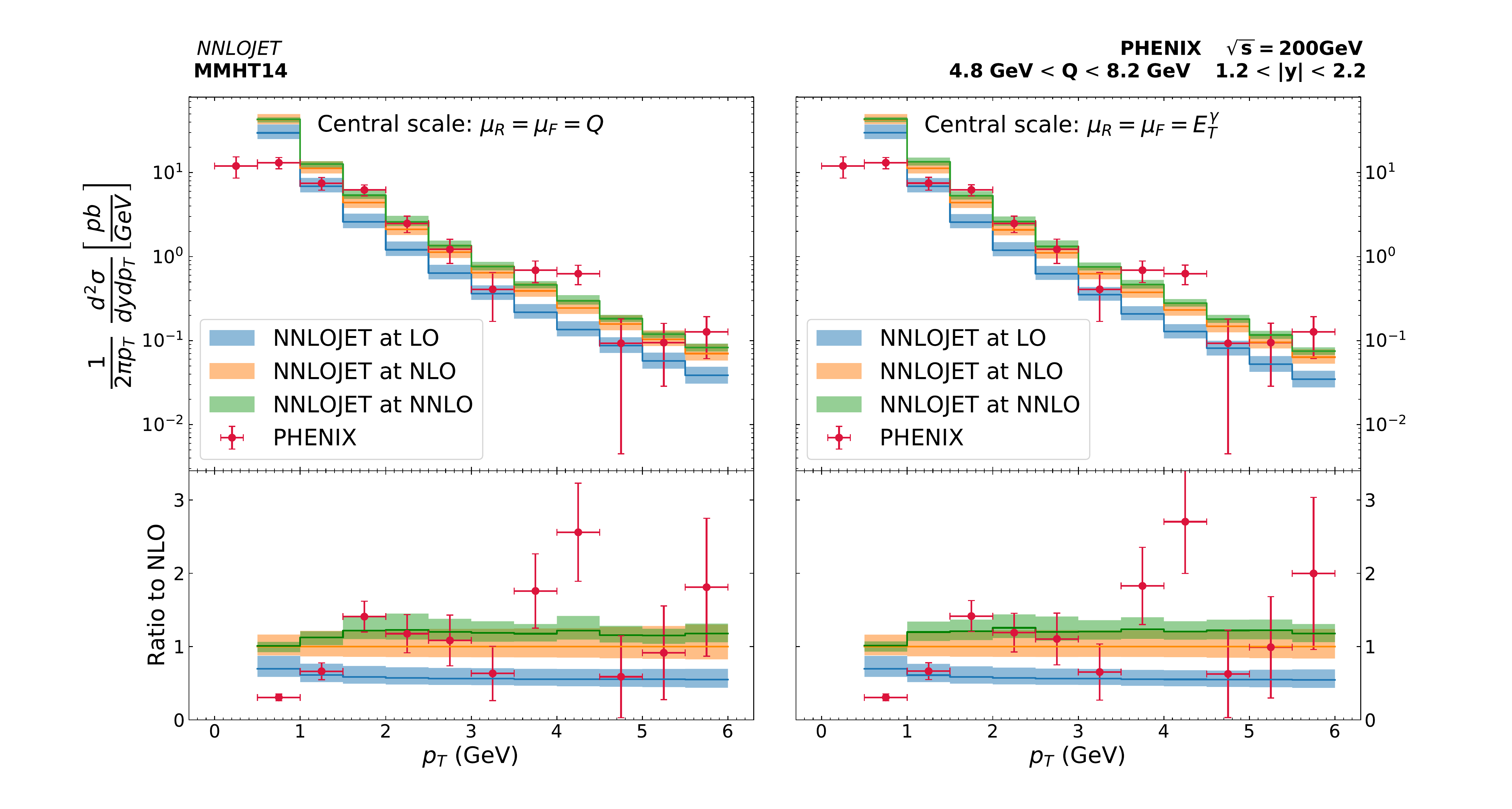}
    \caption{Transverse momentum distribution of lepton pairs, evaluated for 
    central renormalization and factorization scales at $Q$ (left) and $E_T$ (right) 
    and compared to data~\protect\cite{PHENIX:2018dwt} from the PHENIX experiment. }
    \label{fig:phenixbothscales}    
\end{figure}

The predictions for the $p_T$-distribution at LO, NLO and NNLO are displayed in Figure~\ref{fig:phenixbothscales}. 
We consider 
two choices for the central scale used in the evaluations: $\mu_F=\mu_R=Q$ (left) and $\mu_F=\mu_R=E_T$ (right), with $E_T=\sqrt{Q^2+p_T^2}$.
At  values of $p_T>1.5$~GeV, we observe that the NNLO corrections are positive throughout, and almost independent on $p_T$ for both scale choices. The increase over NLO amounts to +25\% for either of the two central scales. 
   \begin{figure}[t]
      \includegraphics[width=\columnwidth]{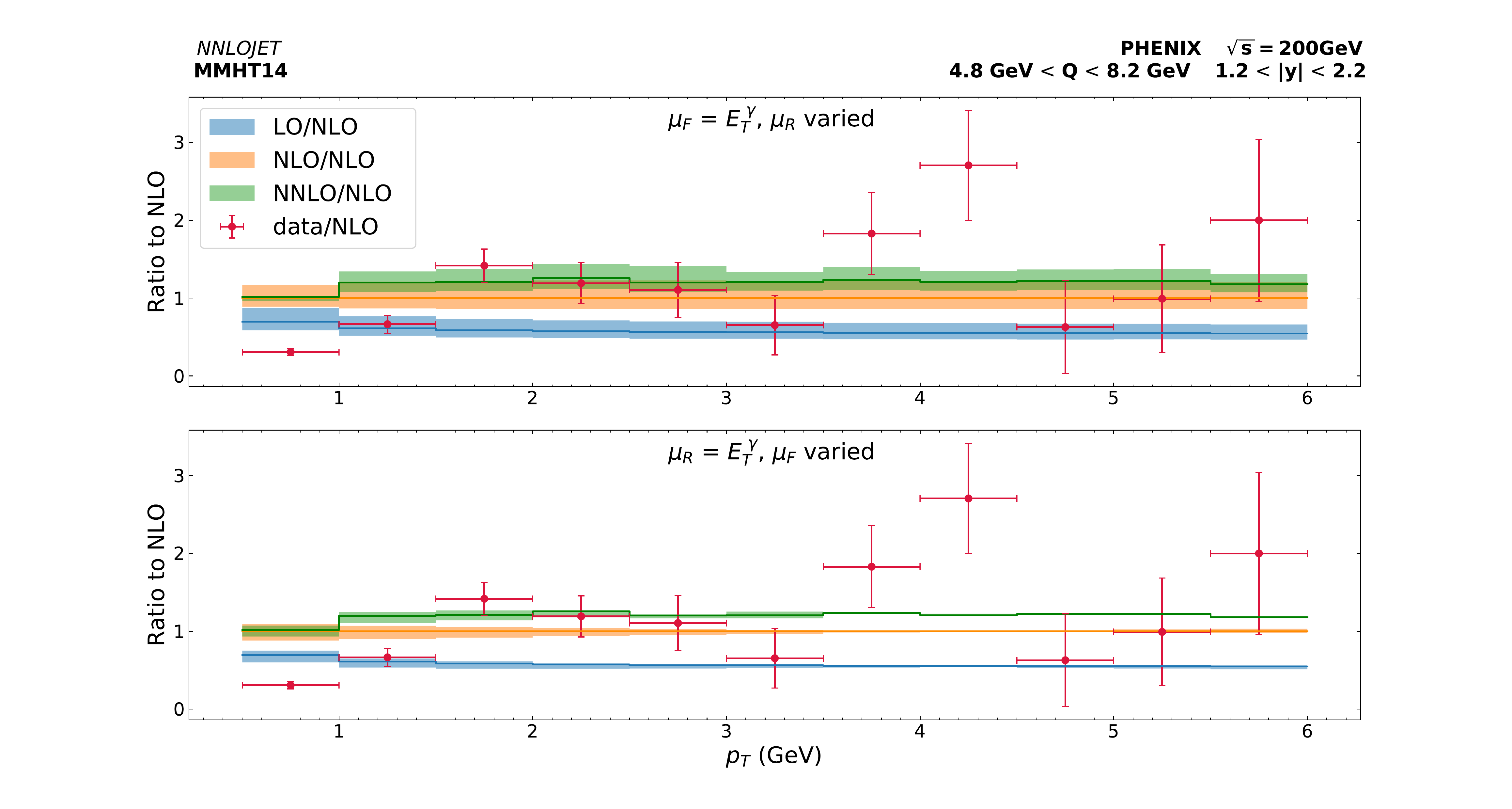}
    \caption{Dependence of transverse momentum distribution on 
    variation of renormalization~(upper frame) and factorization~(lower frame) scales. }
    \label{fig:phenixmufmur}
\end{figure}

The uncertainty on the theory predictions is estimated through an independent variation of $\mu_F$ and $\mu_R$ by a factor 2 around the central scale, 
excluding the two combinations where  $\mu_F$ and $\mu_R$ are changed in opposite directions (seven-point variation). 
We observe that a near-uniform NNLO scale uncertainty of $\pm 15$\% if either $E_T$ or $Q$ is chosen as central scale, and an overlap of the NLO and NNLO 
theory uncertainty bands. Given that the difference between the two choices of central scales is only marginal, $E_T$ is chosen as default central scale for the remainder of this study.  
  The scale dependence arises mainly from the variation of $\mu_R$, as can be seen in Figure~\ref{fig:phenixmufmur}. 
 
 With   $E_T$ as default central scale, we observe from Figure~\ref{fig:phenixbothscales} 
 that the PHENIX data for $p_T>1.5$~GeV are reasonably well-described within their respective uncertainties both at NLO and at NNLO. The inclusion of 
 NNLO corrections leads to a better description of the normalization of the PHENIX data, which is however not significant in view of the 
 large size of the experimental errors. Below $p_T=1.5$~GeV, the fixed-order predictions exceed the data and fail to account for their shape indicating the onset of large logarithmic corrections that requires resummation.
This observed transition into the resummation regime aligns well with a naive estimate based on the leading logarithmic behaviour compensating the suppression of the strong coupling, $C_F \tfrac{\alphas(Q)}{\pi} \ln^2(p_T^2/Q^2) \sim 1$.

\section{Numerical results for NuSea}
\label{sec:results_Nusea}

The NuSea experiment measured Drell-Yan lepton pair production on a fixed target at the Fermilab Tevatron with beam energy of 800 GeV, resulting in 
a centre-of-mass energy of $\sqrt{s} = 38.8$~GeV. Its initial objective was the determination of the flavour decomposition of the  sea quark distributions 
in the proton through the measurement of cross section ratios between proton and deuterium targets~\cite{NuSea:1998kqi}. Subsequent analysis of 
the large NuSea  data set allowed measurements of absolute lepton pair production cross sections on protons~\cite{NuSea:2003qoe,Webb:2003bj}, 
multi-differential in 
the transverse momentum of the pair $p_T$, the invariant mass of the pair $Q$ and Feynman-parameter $x_F= 2p_L/\sqrt{s}$, where $p_L$ is 
the longitudinal momentum of the pair in the center-of-mass frame.  

\begin{figure}[t]
  \includegraphics[width=0.24\columnwidth,page=1]{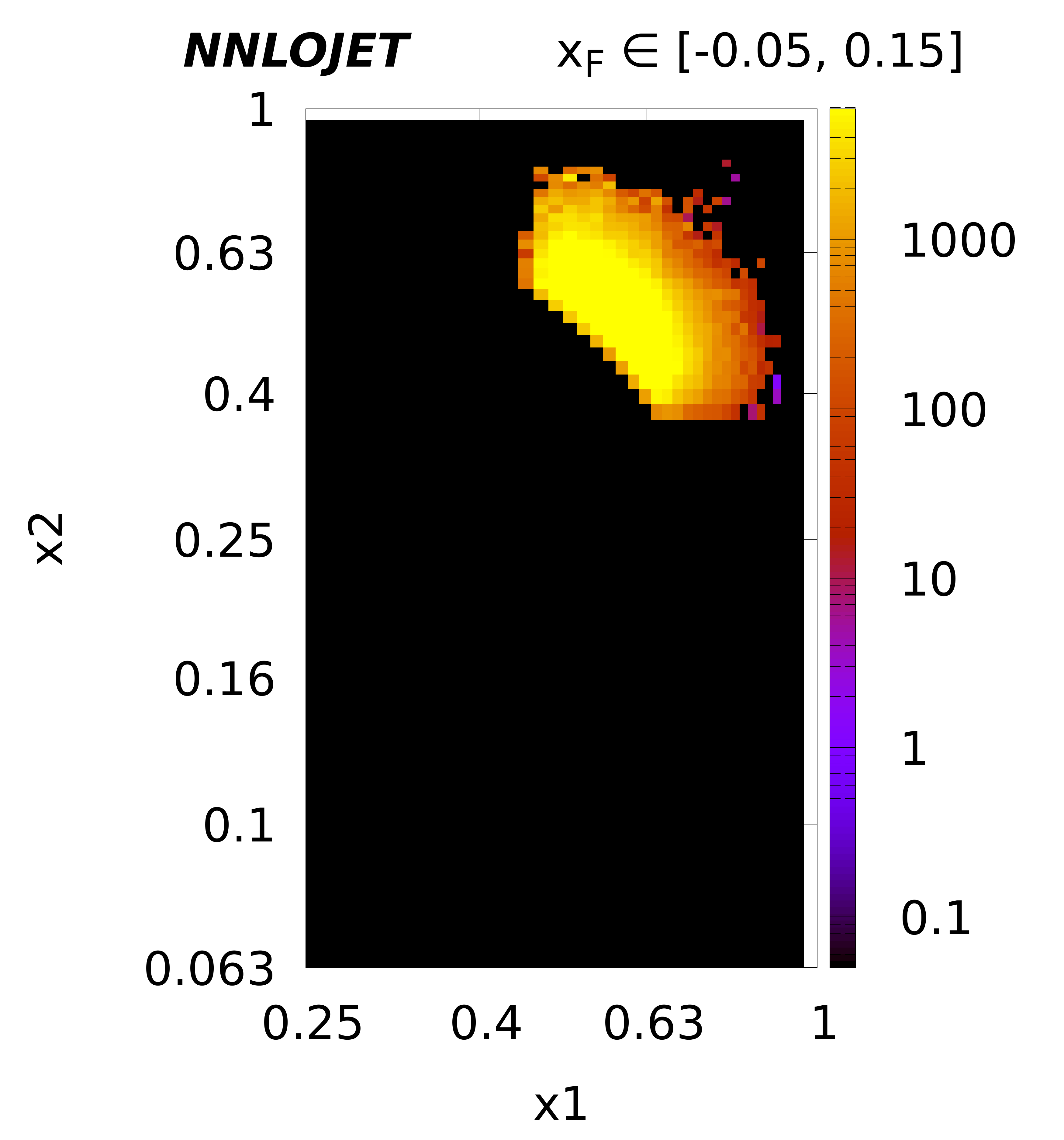}
  \includegraphics[width=0.24\columnwidth,page=2]{Figures/Nusea/x1_x2_MMHT.pdf}
  \includegraphics[width=0.24\columnwidth,page=3]{Figures/Nusea/x1_x2_MMHT.pdf}
  \includegraphics[width=0.24\columnwidth,page=4]{Figures/Nusea/x1_x2_MMHT.pdf}
  \caption{Range of parton momentum fractions $x_1,x_2$ that are resolved by the different $x_F$-bins in the  NuSea measurement~\protect{\cite{Webb:2003bj,NuSea:2003qoe}}, based on a leading-order computation for the mass range
  $4.2$~GeV$<Q<5.2$~GeV. Color indicates the numerical magnitude of the 
  contribution to the cross section.  }
  \label{fig:x1x2}
\end{figure}

These measurements were made in six bins in $Q$, ranging between 4.2~GeV and 16.85~GeV, while excluding the $\Upsilon$ resonances. Each bin in 
$Q$ was subdivided into four bins in $x_F$, between $-0.05$ in the backward direction and $0.8$ in the very forward direction, where the $p_T$ spectrum
was then measured in a range up to 7~GeV at most. Owing to restrictions in the detector acceptance of NuSea, 
the maximal value of $p_T$ is not attained for  all combinations of $Q$ and $x_F$, such that 
the $p_T$ spectrum ends already at lower values in some of the bins. Moreover, especially for the larger values of $Q$ and with $x_F$ in 
the forward region, large values of $p_T$ require parton momentum fractions $(x_1,x_2)$ close to unity at Born-level. This is  illustrated in 
Figure~\ref{fig:x1x2} for  the mass bin $4.2$~GeV$<Q<5.2$~GeV; the higher mass bins 
correspond to even larger values of $x_1$ and $x_2$. In these regions, the fixed-order 
perturbative predictions vanish rapidly due to the decrease of the parton luminosity, while the 
experimental
 measurements 
may turn out to produce sizable cross sections even in these perturbatively disfavoured regions. 

\begin{figure}
    \includegraphics[width=\columnwidth]{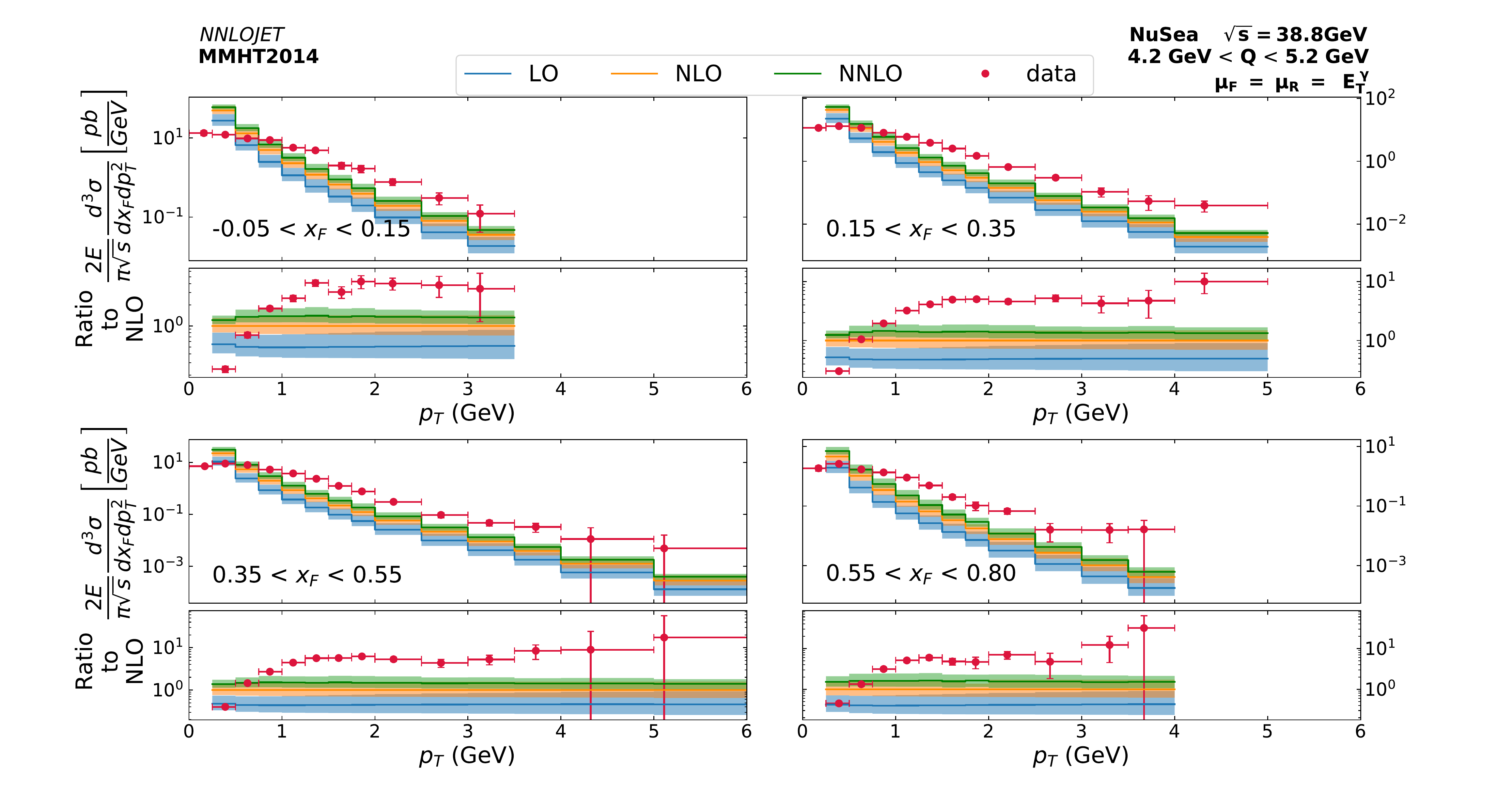}
    \caption{Transverse momentum distribution of lepton pairs with $4.2$~GeV$<Q<5.2$~GeV in different bins in $x_F$, compared to 
    data from the NuSea experiment~\protect{\cite{Webb:2003bj,NuSea:2003qoe}}.}
    \label{fig:nuseaqbin1}
\end{figure}

\begin{figure}
    \includegraphics[width=\columnwidth]{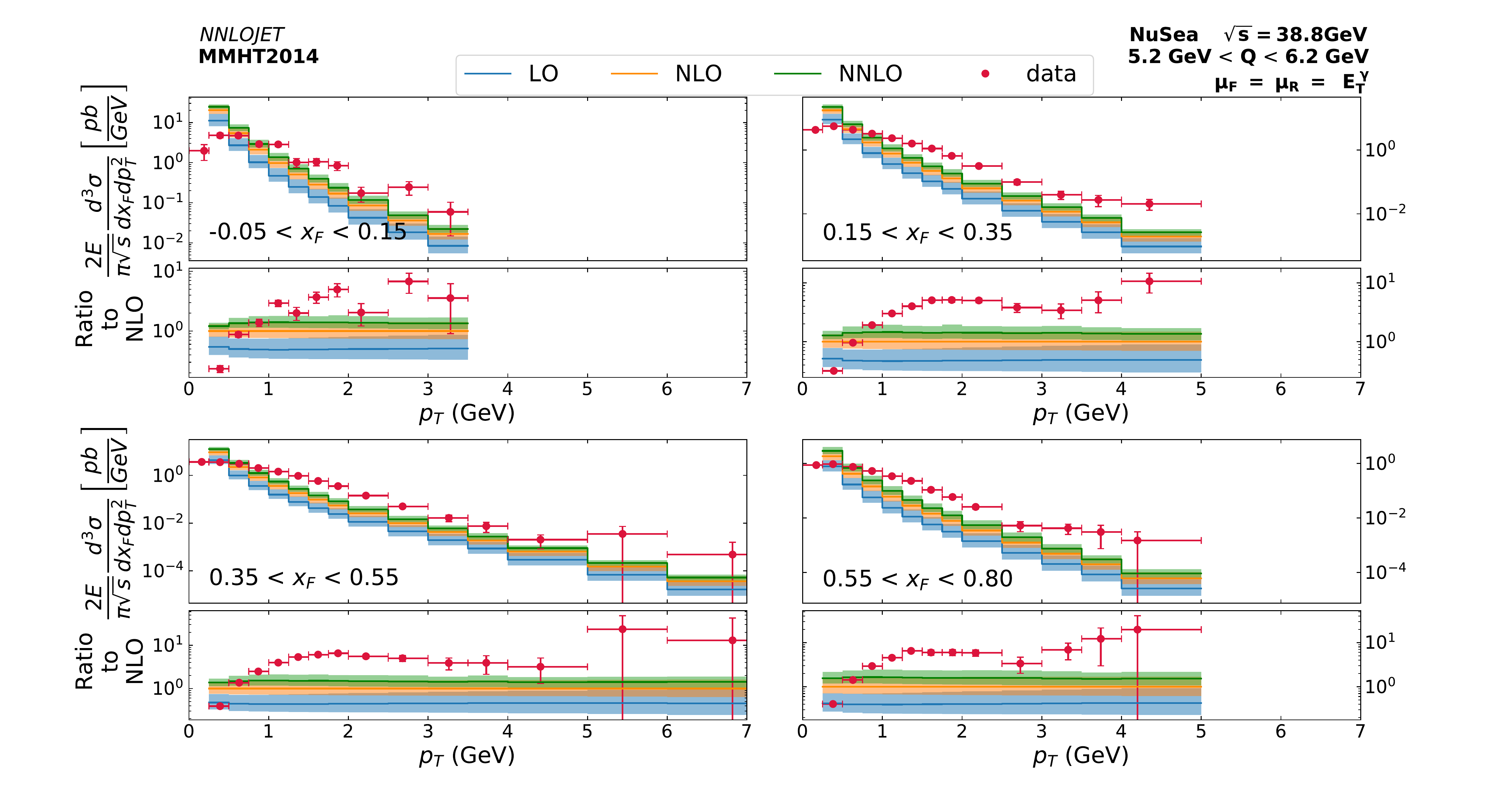}
     \caption{Transverse momentum distribution of lepton pairs with $5.2$~GeV$<Q<6.2$~GeV in different bins in $x_F$, compared to 
    data from the NuSea experiment~\protect{\cite{Webb:2003bj,NuSea:2003qoe}}.}
    \label{fig:nuseaqbin2}
\end{figure}

\begin{figure}
    \includegraphics[width=\columnwidth]{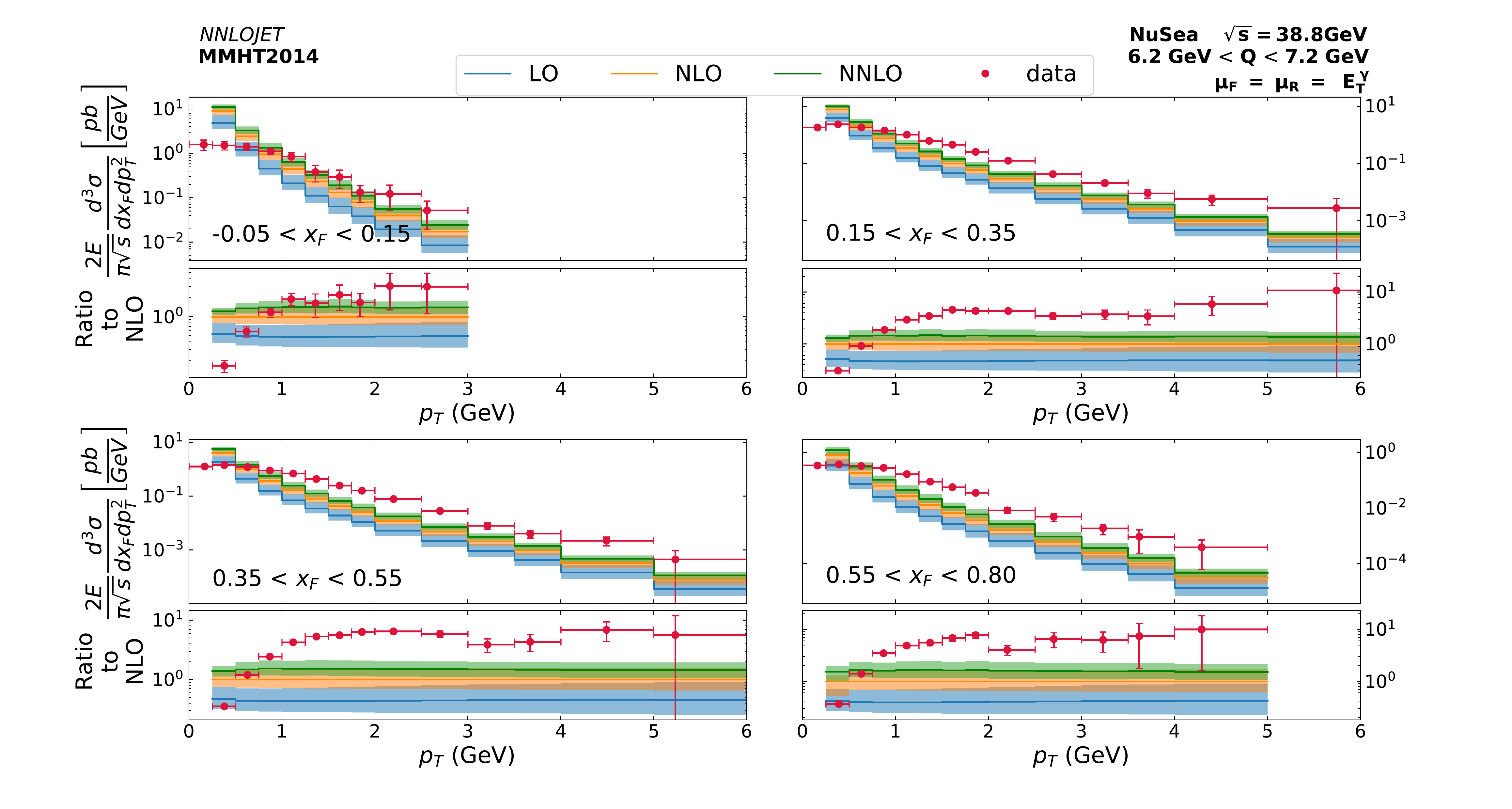}
     \caption{Transverse momentum distribution of lepton pairs with $6.2$~GeV$<Q<7.2$~GeV in different bins in $x_F$, compared to 
    data from the NuSea experiment~\protect{\cite{Webb:2003bj,NuSea:2003qoe}}.}
    \label{fig:nuseaqbin3}
\end{figure}

\begin{figure}
    \includegraphics[width=\columnwidth]{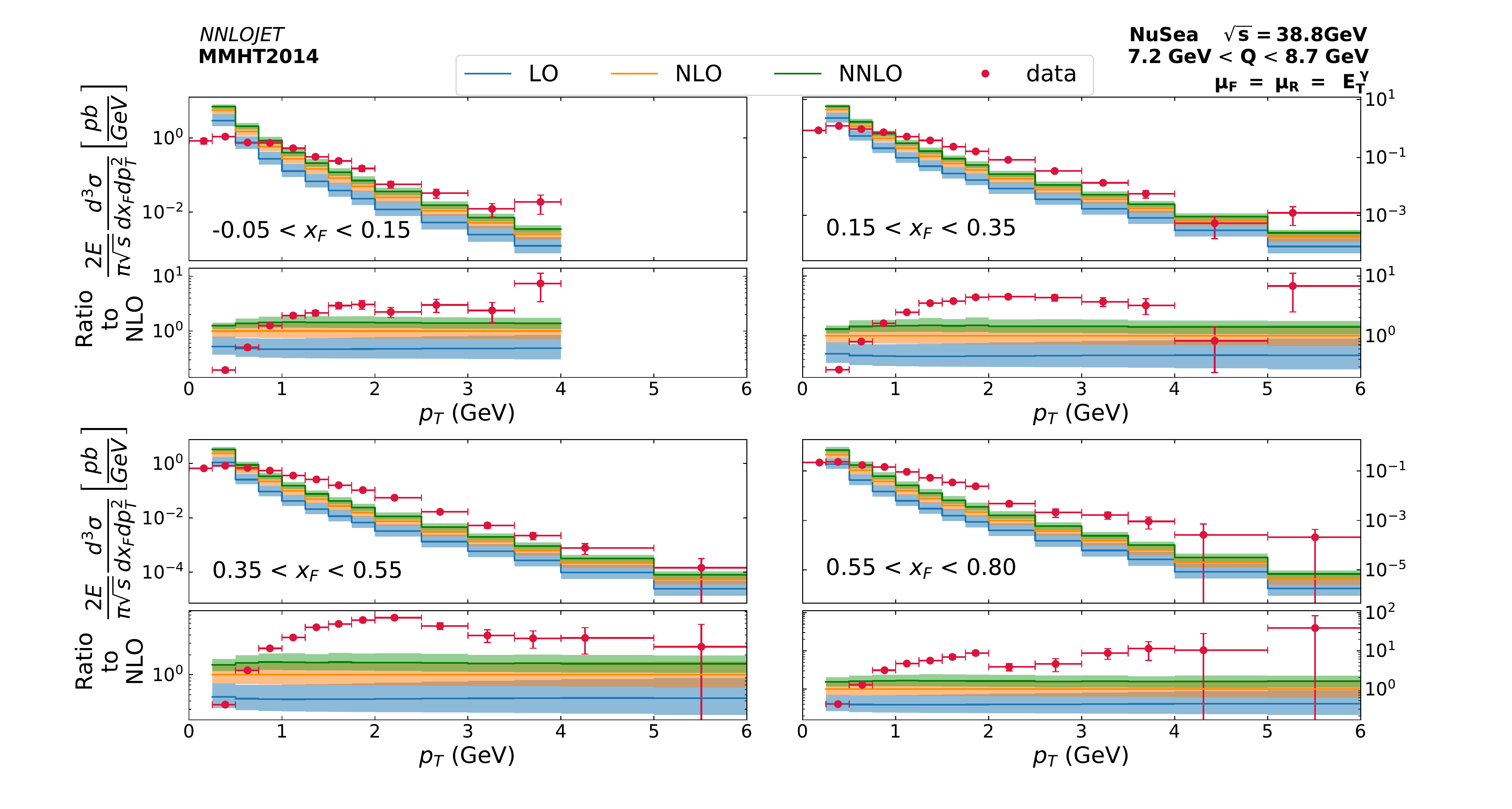}
 \caption{Transverse momentum distribution of lepton pairs with $7.2$~GeV$<Q<8.7$~GeV in different bins in $x_F$, compared to 
    data from the NuSea experiment~\protect{\cite{Webb:2003bj,NuSea:2003qoe}}.}
    \label{fig:nuseaqbin4}
\end{figure}

\begin{figure}
    \includegraphics[width=\columnwidth]{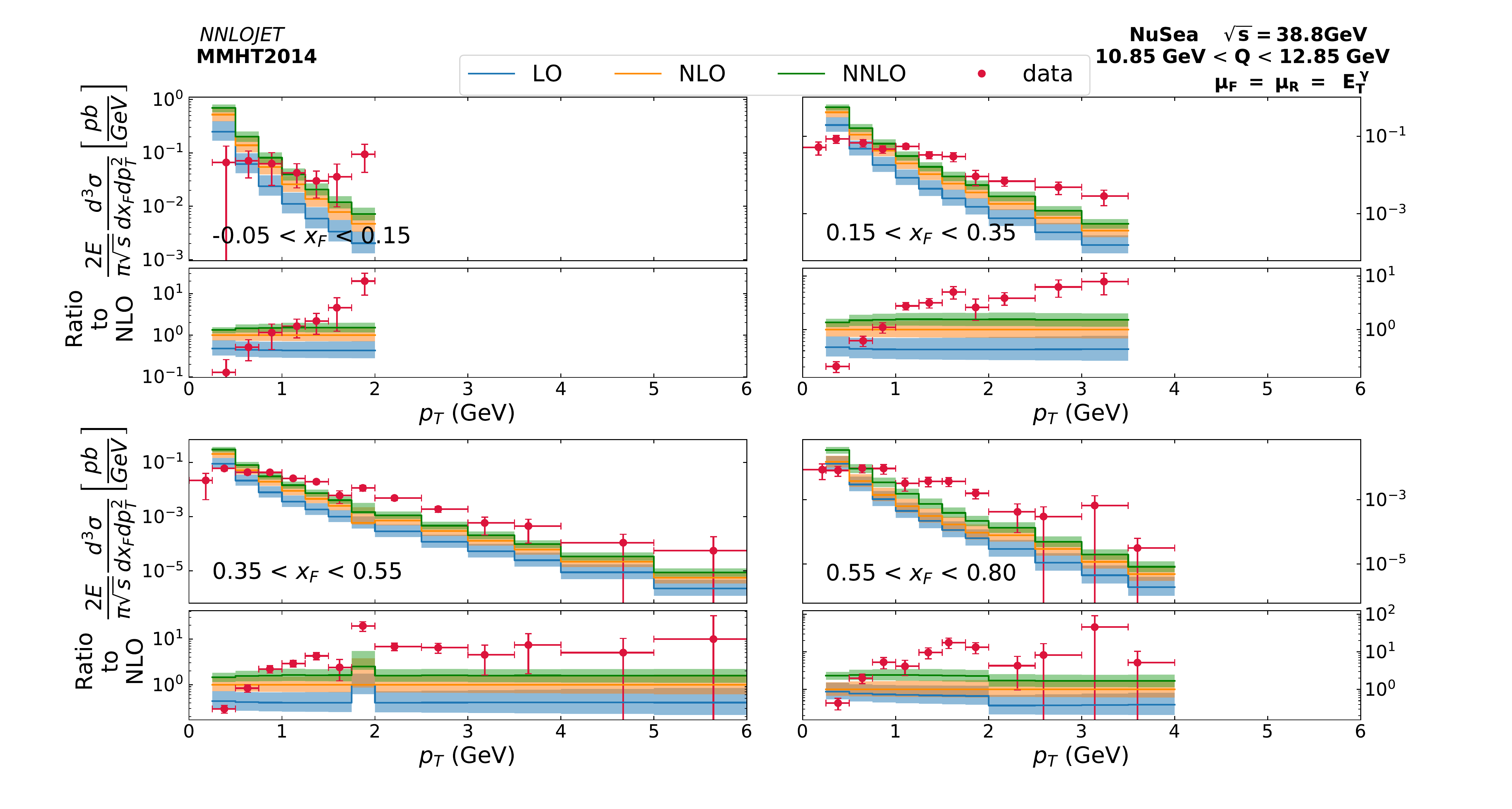}
 \caption{Transverse momentum distribution of lepton pairs with $10.85$~GeV$<Q<12.85$~GeV in different bins in $x_F$, compared to 
    data from the NuSea experiment~\protect{\cite{Webb:2003bj,NuSea:2003qoe}}.}
    \label{fig:nuseaqbin5}
\end{figure}

\begin{figure}
    \includegraphics[width=\columnwidth]{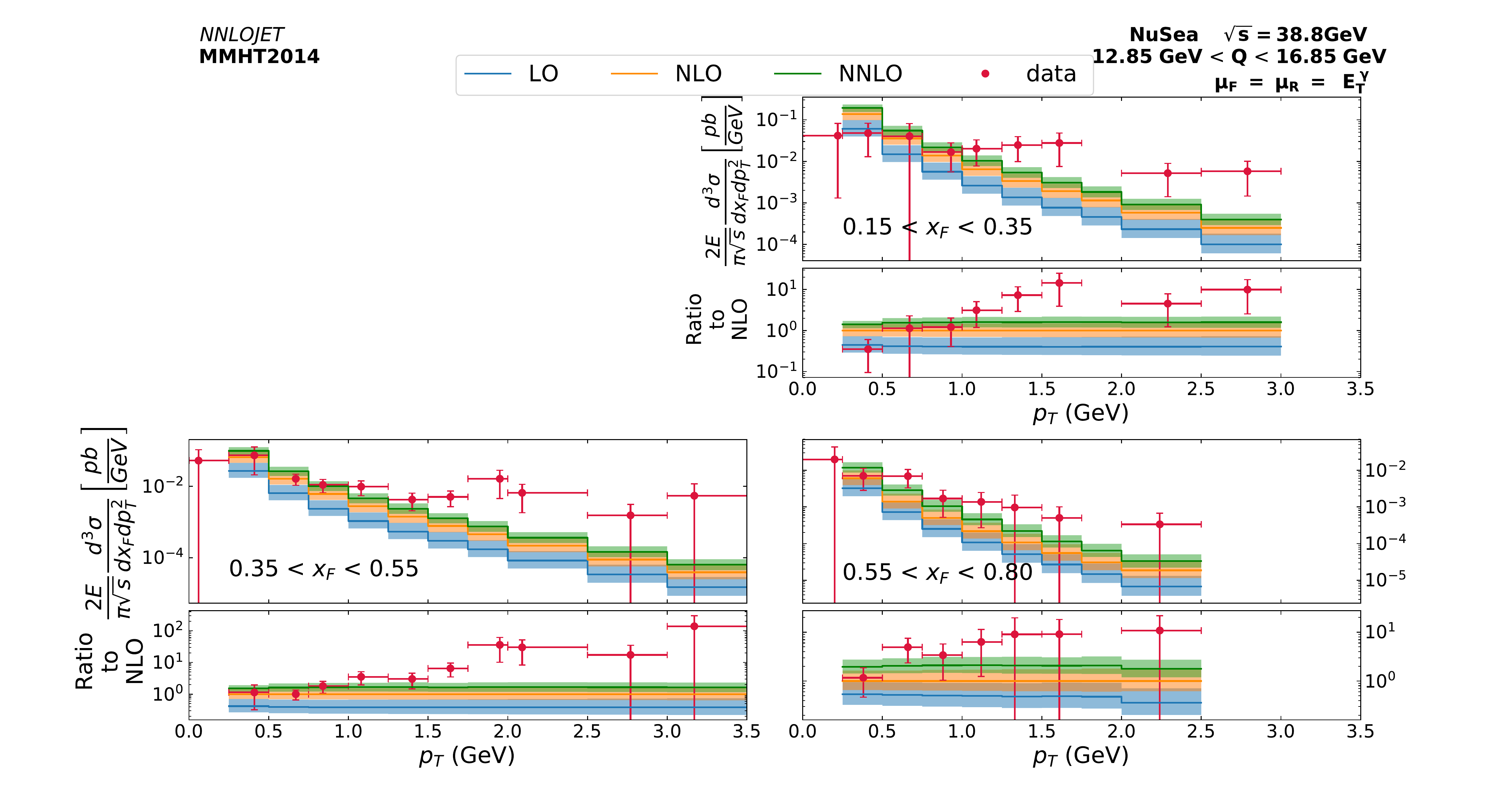}
 \caption{Transverse momentum distribution of lepton pairs with $12.85$~GeV$<Q<16.85$~GeV in different bins in $x_F$, compared to 
    data from the NuSea experiment~\protect{\cite{Webb:2003bj,NuSea:2003qoe}}.}
    \label{fig:nuseaqbin6}
\end{figure}

We compute the $p_T$-distributions for all $(Q,x_F)$ bins where data are available, with \nnlojet using the vector-boson-plus-jet calculation with the jet requirement
replaced by a minimum cut on $p_T$. The results are displayed in Figures~\ref{fig:nuseaqbin1}--\ref{fig:nuseaqbin6}. 
The fixed-order predictions of the $p_T$-distributions diverge for $p_T\to 0$, where an all-order resummation of logarithmically enhanced terms 
is required for a meaningful prediction. Consequently, the left-most bin in $p_T$, which contains the ($p_T=0$)-edge is discarded in our 
fixed-order computation.  
As for the PHENIX data considered in the previous section, we observe that for the NuSea kinematics, central scale choices of $Q$ and $E_T$ yield very similar central cross section values and uncertainty bands. 
We therefore make the choice of $\mu_F=\mu_R=E_T$ for the central scale, and theory uncertainties are estimated through a seven-point variation around the central scale values. 



For several of the $(Q,x_F)$ bins, we observe that the experimental data exceed the theory predictions by one order of magnitude or more. 
 These cases typically correspond 
to kinematical situations that require one of the parton momentum fractions  to be very close to unity. The large excess  
of experimental cross section measurements over theory expectations
in these kinematical ranges may hint to large non-perturbative effects that 
distort the $p_T$ spectra. 

We observe that the NNLO QCD corrections are largely $p_T$-independent in all $(Q,x_F)$ bins. They typically amount to an increase over 
the previously known NLO results which depend on the $Q$ bin as follows: $+35\%$ for the lower two bins (Figures \ref{fig:nuseaqbin1} and \ref{fig:nuseaqbin2}), $+50\%$ for the next two (Figures \ref{fig:nuseaqbin3} and \ref{fig:nuseaqbin4}) and over +70\% for the two bins of highest $Q$ (Figures \ref{fig:nuseaqbin5} and \ref{fig:nuseaqbin6}). 
A similar pattern is increasing corrections was already observed in going from LO to NLO. The  increase at NNLO is within the 
NLO scale uncertainty bands for all $Q$ bins. The scale uncertainties are reduced from NLO to NNLO,  and are typically in a range around $\pm 40\%$. 
The size of the NNLO corrections and the associated magnitude of the NNLO theory 
uncertainty is larger than in the case of PHENIX. This feature may be related to the lower centre-of-mass energy  at NuSea, which 
consequently probes large parton momentum fractions, see Figure~\ref{fig:x1x2}, where threshold logarithms are starting to influence the 
behaviour of the perturbative higher-order coefficients starting from NLO onwards. 
The onset of threshold logarithms at $(x_1,x_2)\to 1$ also explains the increase of the corrections towards larger $Q$. 

The newly computed NNLO corrections do not explain the large discrepancy between data and theory that was also previously observed at 
NLO level~\cite{Bacchetta:2019tcu,BermudezMartinez:2020tys}. 
In the majority of the $(Q,x_F)$ bins, the ratio between data and NNLO theory approaches a constant value 
for $p_T\gapprox 1$~GeV. The values of this ratio range between 1.5 for the central $x_F$ bins and up to 10 for the forward $x_F$ bins, 
with only a marginal dependence on the invariant mass $Q$. For $p_T<1$~GeV, we do not expect the fixed order predictions to be meaningful due to their divergent behaviour at low $p_T$. 

For increasing $x_F$, we observe that the $p_T$-distributions fall considerably more steeply towards larger values of $p_T$. This steeper decrease 
directly correlates with the larger data/theory ratios. It is very suggestive to observe that the $p_T$ distributions in the vast majority of 
the $(Q,x_F)$ bins could be brought in agreement with the data by shifting the value of $p_T$ in the theory predictions 
by about $+1$~GeV when comparing to data. This observation could point to an intrinsic transverse momentum of the partons in the proton, which is caused by their non-perturbative bound-state dynamics.

Earlier studies have modelled this intrinsic transverse momentum effect in detail.
In~\cite{BermudezMartinez:2020tys}, it was shown that using transverse momentum dependent parton distributions (TMD)
in combination with NLO fixed-order theory matched to parton-shower in the MC@NLO framework~\cite{Alwall:2014hca}
leads to a satisfactory description of the NuSea data for the first bin in $x_F$ and $Q$: ($-0.05<x_F<0.15$ ; $4.2$~GeV$<Q<5.2$~GeV).
This study could be repeated for the full range of $(Q,x_F)$ in the light of the newly computed NNLO corrections.

The newly computed NNLO corrections are positive throughout all bins in $x_F$ and $Q$, and largely $p_T$-independent. They indicate the need for a somewhat smaller shift of the $p_T$ spectrum, as compared to NLO. This feature can be understood from the 
extra partonic recoil from one additional emission 
between NLO and NNLO, which does account for some (albeit small) part of the  intrinsic transverse 
momentum. However, our results clearly show that perturbative emissions are insufficient to explain 
the $p_T$-spectrum at fixed target energies.

PDF uncertainties within the MMHT14 set were studied and found to be smaller than the residual scale uncertainties.
It should however be emphasised that the behaviour of PDFs in the region relevant to the NuSea measurements is only poorly constrained 
by experimental data, such that their behaviour is largely determined by the extrapolation of the functional forms used in the PDF fitting procedure 
to large values of $x$.   
We have also re-computed the NNLO predictions for NuSea with the NNPDF3.1 parton distribution functions~\cite{NNPDF:2017mvq}. In contrast 
to MMHT14, the NNPDF3.1 quark and anti-quark distributions are negative at large values of $x$, which results in negative cross section 
predictions with the NNPDF3.1 set in several bins in $x_F$, especially for the larger $Q$ values.
Since the release of NNPDF3.1, positivity of parton distributions at large $x$ has been revisited~\cite{Candido:2020yat}, it is now an inherent constraint in the new NNPDF4.0 release~\cite{Ball:2021leu}, while still remaining a matter of ongoing debate on formal grounds~\cite{Collins:2021vke}.

The NuSea data could provide important and unique constraints to the quark distributions in the proton at very large $x$, a region not covered by any other measurement. For them to be included in a global PDF determination will however require substantial advances in the understanding and quantitative modelling of non-perturbative effects that distort the $p_T$  spectra in the NuSea kinematical range.

\section{Conclusions}

In this letter, we have investigated the impact of NNLO QCD corrections on the $p_T$-distribution of low-mass lepton pairs. This process is particularly 
interesting in view of probing the transition region between perturbative and non-perturbative dynamics in QCD. Previous studies~\cite{Bacchetta:2019tcu,BermudezMartinez:2020tys} based 
on NLO QCD have highlighted a poor perturbative convergence and displayed substantial discrepancies between theory predictions 
and experimental data. We focused our study on two representative 
data sets from PHENIX ($\sqrt{s}=200$~GeV) and NuSea ($\sqrt{s}=38.8$~GeV), which probe similar final-state kinematics at different collision 
energies, thereby probing different regions of parton momentum fractions.
In either case, we observe large positive NNLO QCD corrections in the range of $+25\%$ in the case of the theoretical predictions to the PHENIX data and between $+35\%$ and $+70\%$ in the case of the theoretical predictions to the NuSea data, which are within the previously quoted NLO theory 
uncertainties, and which indicate the onset of perturbative convergence. The PHENIX data are described at NNLO QCD in a satisfactory manner. 
In contrast, the NNLO QCD prediction remains considerably below the NuSea data. This discrepancy increases with increasing $x_F$, which translates into larger values of parton momentum fractions being probed, which may indicate enhanced sensitivity to non-perturbative effects in these extreme kinematical regions.

\acknowledgments

This research is supported by the Dutch Organisation for Scientific Research (NWO) through the VENI grant 680-47-461,  
by the UK Science and Technology Facilities Council (STFC) through grant ST/T001011/1, 
by the Swiss National Science Foundation (SNF) under contract 200021-197130 
and 
by the Swiss National Supercomputing Centre (CSCS) under project ID ETH5f.

Furthermore we thank Xuan Chen, Juan Cruz-Martinez, James Currie,  Marius Höfer,  Matteo Marcoli, Jonathan Mo, Tom Morgan, Jan Niehues, Robin Sch\"urmann, João Pires, Duncan Walker, and James Whitehead for useful discussions and their many contributions to the \nnlojet code.

\bibliography{DY}

\end{document}